\documentclass[a4paper]{jpconf}
\usepackage{graphicx}
\begin{document}
\title{Search for gravitational waves associated with the InterPlanetary Network short gamma ray bursts}

\author{V. Predoi for LIGO Scientific Collaboration and Virgo Collaboration and Kevin Hurley for the IPN }

\address{Cardiff University, Physics and Astronomy, The Parade, Cardiff CF24 3AA, UK}

\ead{valeriu.predoi@astro.cf.ac.uk}

\begin{abstract}
We outline the scientific motivation behind a search for gravitational waves associated with short gamma ray bursts detected by the InterPlanetary Network (IPN) during LIGO's fifth science run and Virgo's first science run. The IPN localisation of short gamma ray bursts is limited to extended error boxes of different shapes and sizes and a search on these error boxes poses a series of challenges for data analysis. We will discuss these challenges and outline the methods to optimise the search over these error boxes.
\end{abstract}

\section{Introduction}

Short hard gamma ray bursts (short GRBs) are believed to be produced by mergers of either two neutron stars or a neutron star and a stellar mass black hole \cite{Nakar:2007, ShibTan06, Berger:2010qx}. These events are ideal sources for strong gravitational wave emission \cite{ACST94, Kiuchi:2010ze}. If an observation of both gamma rays and gravitational waves (GW) originating from the same event could be achieved, it will increase confidence and allow for better science output. It is thus of great importance to constantly monitor and record GRBs to allow a GW search to be performed around the burst times. Systematic analyses of GW data around short GRB times have been done in the past and the most recent publications from the LIGO-Virgo group contain results from the Swift-observed GRBs during LIGO's fifth science run (S5) and Virgo's first science run (VSR1) \cite{Abadie:2010uf, Collaboration:2009kk}. This paper reports the methodology and motivation for a proposed GW search around the times of 20 additional GRB during S5/VSR1. These bursts were observed by the InterPlanetary Network (IPN) \cite{Hurley:2002wv, HurleyHTML}, a group of satellites orbiting the Earth and Mars and operating, among other equipment on board, gamma ray detectors. These bursts were detected between 2006 and 2007 and have been localized, in both time and sky location, such that a targeted GW search is possible.

The InterPlanetary Network \cite{Hurley:2002wv, Hurley:1999ym} employs several space missions and synthesizes data obtained from the detection of the same burst by different spacecraft equipped with gamma ray detectors. The IPN has been operating for three successive generations; presently the third IPN (IPN3) began its operation in November 1990. Currently the spacecraft gathering data are Konus-WIND, Suzaku, INTEGRAL,  RHESSI, Swift, Fermi/GBM (in Earth orbit), MESSENGER (in Mercury orbit) and Mars Odyssey (in Mars orbit) \cite{HurleyHTML}. When the duty cycles and effective fields of view of all the missions in the network are considered, the IPN is an all-time, isotropic GRB monitor.

In this paper, we first review the motivation for performing a joint GW-GRB search as an overview and then provide details of the IPN short GRBs that were identified during the LIGO-Virgo S5-VSR1 science runs.  We discuss the necessary changes to the current ongoing analysis (for the S6/VSR2-3 science runs Swift and Fermi-observed GRBs, \cite{lvc:s6grb, Harry:2010fr}) that are needed to implement a search on GRBs identified by the IPN network which may be less well localized in both sky position and time than the corresponding bursts identified by the Swift satellite and used in previous analyses \cite{Abadie:2010uf, Collaboration:2009kk}.

\section{Search for GW associated with short GRB}

Gamma ray bursts fall into two commonly accepted groups depending on their duration and spectral hardness \cite{Nakar:2007, Qinx:2010kp}. Characteristic duration is quantitatively expressed by the $T_{90}$ parameter, defined as the time interval over which 90\% of the total background-subtracted counts are observed (the total counts emitted by the source, found from the counts in the source region minus the contribution from the background), with the interval starting when 5\% of the total counts have been observed \cite{McBreen:2001fd}.  The majority of bursts, with softer spectrum and duration $T_{90} > \mathrm{2s}$ are called long GRBs. They are thought to be produced during a core-collapse supernova event.  Those with a harder spectrum and $T_{90} \leq \mathrm{2s}$ are called short GRBs.  

Short GRBs ($T_{90} \leq \mathrm{2s}$) are widely believed to be mergers of either neutron star-neutron star or neutron star-black hole binaries \cite{Nakar:2007}. Such compact binary coalescences are believed to generate strong gravitational waves in the sensitive frequency band of Earth-based gravitational-wave detectors \cite{Blanchet:2001aw,BD89}. Although binary coalescence is the favoured progenitor model for short GRBs, this has not yet been confirmed by means of a direct observation.  Consequently, the detection of gravitational waves associated with a short GRB would provide evidence that the progenitor is indeed a coalescing compact binary and also provide information on the parameters of the binary.  

The coalescence of a compact stellar-mass binary (either two neutron stars or a neutron star and a black hole companion) is the endpoint of its $\sim$Gyr life revolving around the common centre of mass while constantly losing energy and angular momentum through emission of gravitational radiation. The binary merger will lead to the formation of either a transient hyper massive highly magnetized neutron star with a lifetime of a few ms \cite{Shibata:2005mz, Duez:2005cj} that will collapse to a rapidly spinning black hole or straight to the formation of a black hole \cite{ShibTan06, Shibata:2007zm}.  In the favoured model of short GRBs \cite{Shibata:2005mz, Kiuchi:2010ze, Rezzolla:2011da, Oechslin:2005mw}, the gamma-ray emission is contingent on the formation of a massive (and highly magnetized) torus around the final black hole. The matter in the torus is accelerated to relativistic velocities leading to the formation of a collimated jet of electromagnetic radiation along the axis of the former binary total angular momentum. Differences in velocities between layers in the jet account for the prompt gamma-ray emission. Full relativistic and magnetohydrodynamic numerical simulations have shown that the time difference between the binary merger and the jet formation can be between a few milliseconds up to a few seconds \cite{ShibTan06, Shibata:2007zm, Rezzolla:2011da, Oechslin:2005mw}. 

Supposing gravitational waves produced just before merger travel at the same speed as the gamma rays, the speed of light in vacuum, and suppose the observer is situated within the cone of the jet, one would expect to observe the GW within a few seconds prior to the arrival of the gamma rays.
 As the GWs emitted during the pre-merger inspiral dominate the signal-to-noise-ratio (SNR) observable in current GW detector data and they can be accurately modelled using post-Newtonian approximations \cite{ABIQ04, AIQS06a}, our search will be aimed at this phase only.

\section{The short GRB sample to be analysed}

We gathered the sample of GRBs to be analysed using data provided in the IPN online table available publicly at \cite{HurleyHTML} and by cross-checking this with NASA's HEASARC online table at \cite{heasarc}. Since these GRBs are not reported in any of NASA's Gamma Ray Burst Circulars (GCN), a manual check on each burst had to be performed in order to confirm its characteristics, involving checking several IPN satellite homepages e.g. Suzaku at \cite{suzaku}, INTEGRAL at \cite{integral}, Swift at \cite{swift} and Konus-WIND at \cite{konus}.

The details of the sky position were obtained by manually processing raw data from the IPN satellites for every GRB.  In order to perform a search of the GW data for a given GRB, the sky position and time of the GRB must be determined.  This information is obtained from knowledge of: the IPN satellites that detected the burst together with their absolute and relative sky positions (information needed for constructing the GRB error boxes), locations of all the spacecraft (used to obtain the blocking constraints to reduce the size of the GRB error boxes), the burst time of arrival at the satellites and at Earth (used to determine the time interval on which we will perform the GW search) and its error. Constructing the GRB error boxes, determining their sizes and choosing the right GW data for analysis are entirely contingent on these pieces of information.

\begin{figure}[htb]
\begin{center}
\includegraphics[height=15pc]{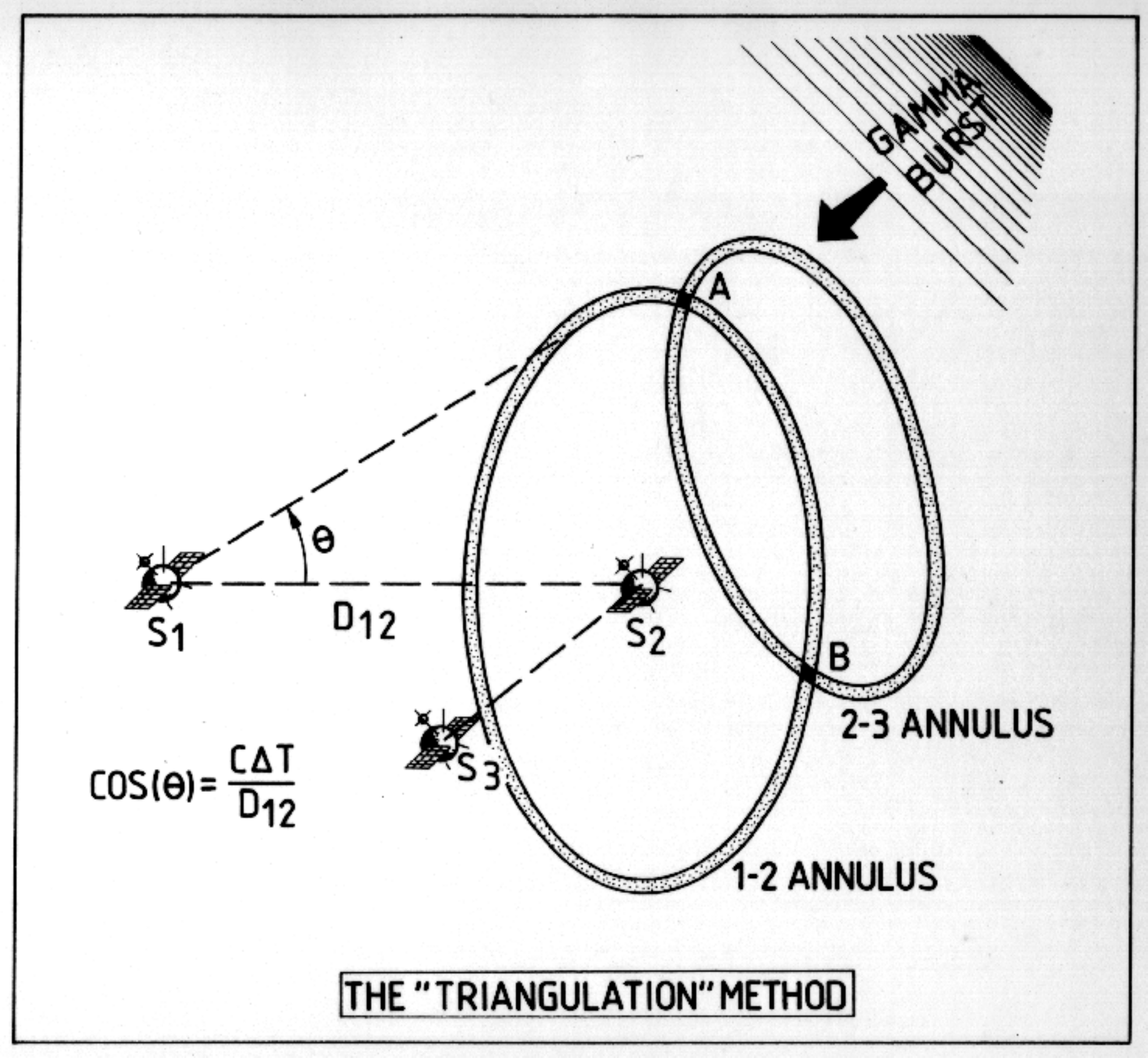}
\caption{\label{fig:triangulation}The IPN schematics: triangulating the position of a GRB using three IPN spacecraft (S1, S2 and S3). Using three satellites we obtain two IPN annuli that intersect to form two error boxes. Reference \cite{HurleyHTML}}
\end{center}
\label{IPNtriangulation}
\end{figure}

\subsection{The IPN triangulation mechanism}
The principle on which the IPN triangulation method is based uses the arrival time of the same burst at different spacecraft to determine the source sky location. Figure \ref{fig:triangulation} illustrates how an IPN triangulation works. S1, S2 and S3 denote three spacecraft detecting the same GRB and $\theta$ is angle between the burst direction and the baseline between S1 and S2.  Then, the burst will be detected by S2 at a time $\delta T$ seconds earlier than S1 
\begin{equation}
\cos (\theta) = c \delta T / D_{12}
\label{eqn1}
\end{equation}
where $D_{12}$ is the distance between S1 and S2, and $c$ the speed of light. Since $D_{12}$ is known and $\delta T$ is measured, $\theta$ is estimated. The solution to equation (1) is represented by a ring, or an $annulus$, whose width depends on the timing uncertainties ($\sigma(\delta T)$) and on the separation $D_{12}$.  The farther apart the detectors, the more precise the localisation. The number of independent couples of detectors (and, therefore, the number of independent annuli) is two for the case of three spacecraft; thus, the burst direction must be inside one of the two intersection regions of the annuli. This intersection region is called the IPN error box of the GRB. Depending on the location of the IPN spacecraft and their timing errors this error box may vary in size from fractions of, to hundreds of square degrees. The annulus width is obtained by propagating the uncertainty on the time delay $\delta T$. Thus, from equation (1) it follows that
\begin{equation}
\sigma(\theta) = \frac{c\sigma (\delta T)}{D_{12} \sin (\theta)}
\label{eqn2}
\end{equation}

Equation (\ref{eqn2}) gives the uncertainty $\sigma(\theta)$ in the angle $\theta$ expressed in radians.  The uncertainty on $D_{12}$ has been neglected, as the main contribution comes from the timing uncertainties. One has to take into account not only the time resolution of each of the detectors, but also the difficulty of comparing different light curves, often derived from different energy bands. For example, when $D_{12}$ spans the typical range: few $10^2$ - few $10^3$ light seconds, then from equation (\ref{eqn2}) it comes out that a minimum time resolution of the order of $10^{-2} - 10^{-3}$ s is required, in order to have $\sigma(\theta) \sim$ few arcminutes or less in order to obtain a precise localisation.

\begin{figure}[htb]
\begin{center}
\includegraphics[width=32pc]{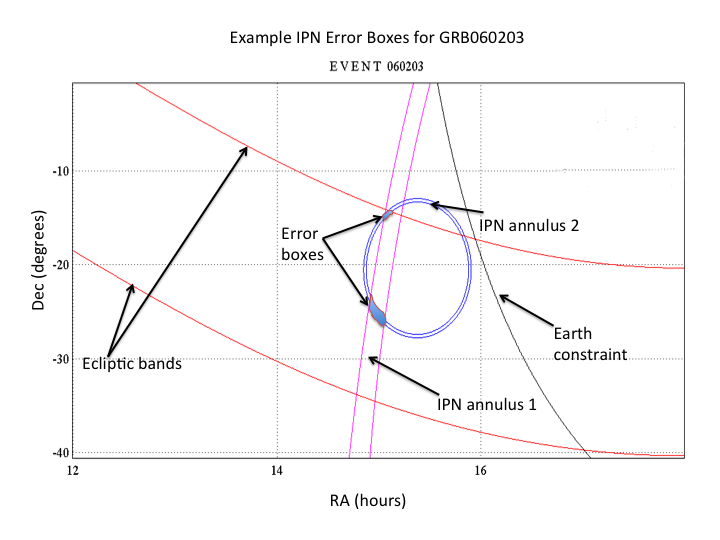}
\caption{\label{fig:error_box}Error box construction: red lines correspond to the Konus-WIND ecliptic latitude bands, blue lines are the 3-$\sigma$ IPN1-2 annulus and magenta lines are the 3-$\sigma$ IPN1-3 annulus obtained by triangulation (in this case we have three IPN spacecraft observing the same burst). The error boxes are the solid regions bounded by the intersections of the IPN annuli and ecliptic bands. }
\end{center}
\label{errorbox}
\end{figure}

\subsection{The IPN GRB error box construction}

The error boxes for the IPN GRBs are constructed from the intersection of the triangulated 3-$\sigma$ IPN annuli and different field-of-view blocking constraints (if present). See figure \ref{fig:error_box} for an illustration example. When the constraints involve Konus-WIND ecliptic latitude bands, the error box that will be kept is located between the ecliptic bands. The reason for this is that the Konus-WIND spacecraft consists of two detectors, one facing the north ecliptic pole, and the other facing the south ecliptic pole. By comparing the count rates on these two detectors, the Konus team obtains an estimate of the ecliptic latitude of the burst. Typically, the band is 20-30 degrees wide, and it is good to a 90-95\% confidence; systematic and statistical errors usually prevent it from being much more accurate than this value. In those cases where we have a single triangulation annulus, plus an ecliptic latitude band, the result is often two long, narrow error boxes, where the annulus intersects the band. Other intersections are possible, too, such as a single long, narrow error box if the IPN annulus grazes the ecliptic latitude band. Also, anything that overlaps a region where a planet blocks the view will be discarded.

\subsection{Determining the GRB time of arrival}

The Earth crossing time, also referred to as the time of arrival of the burst, is the time when the gamma ray signal would cross the centre of the Earth and is the reference time to be considered when constructing the time search window for gravitational waves. 
This time can only be estimated based on the burst arrival times at the different IPN satellites and based on their positions with respect to Earth. This way of choosing the burst time is prone to two types of uncertainties: the first is simply the fact that we may not have the time at Earth but at a satellite that is separated from Earth by a certain distance; the error is directly proportional to the distance to Earth where the closest IPN satellite is located at the time of the burst. The best estimate comes from any satellite that is not interplanetary (i.e. not on orbit around Mars or at a distant point from Earth).  These ``close'' satellites are usually within a fraction of a light second distance from Earth.  These uncertainties are small, less than 1 second for satellites orbiting Earth but may be of up to 5 seconds or more for interplanetary satellites. The second kind of uncertainty comes from the fact that the IPN consists of nine spacecraft with different energy ranges and different spectral sensitivities. So it is possible, in an extreme case, to have one spacecraft trigger on a GRB precursor, while the others trigger on the main burst, resulting in a time difference.  This way, the trigger times can differ by tens of seconds or more.  For short GRBs this effect is minimised due to the hard nature of their spectra and consequently reduces to under one second imprecisions. Altogether, time imprecisions are not more than 5s for the burst sample we will be analysing. The few GRBs that have a timing imprecision greater than 1 second are localised by distant satellites, usually MESSENGER/Mars Odyssey and/or Konus-WIND.

\subsection{Gravitational wave detectors and data availability}
Our aim is to perform a search for GW associated with the well or fairly well localised short GRBs detected by the IPN during LIGO's S5 run that lasted from 4 November 2005 to 30 September 2007, and Virgo's VSR1 that lasted from 18 May 2007 to 30 September 2007. The analysis will make use of data from four operational GW interferometers: the 4km and 2km co-located LIGO detectors at Hanford, WA (H1 and H2), the 4km detector at Livingston, LA (L1) and the 4km detector at Cascina, Italy (V1) \cite{Abbott:2007kv, Abadie:2010px, virgostatus}. The search will be using a method that coherently combines data from multiple operational GW detectors described in \cite{Harry:2010fr}; the method is being used in GW-GRB searches for S6/VSR2-3 \cite{lvc:s6grb} and is proven to be performing better than the method used for the S5/VSR1 search \cite{Harry:2010fr, Abadie:2010uf, Collaboration:2009kk}; we require that all GRBs have data from at least two GW detectors. In order to perform the search, we require approximately forty minutes of data around the time of the GRB. The search pipeline identifies a foreground time representing the time interval around the actual burst when the signal is most likely to have been detected by the GW detectors. For the IPN GRBs that have a burst time of arrival error less than a second, based on the delay between the time of the arrival of the gamma ray and of the GW (described above) an interval of 5 seconds prior to the GRB time and 1 second following it will be used as foreground. This time interval will account for any timing imprecisions and has been previously used in the S5/VSR1 search for GW associated with the Swift GRBs \cite{Abadie:2010uf}. For the IPN GRBs that have a time of arrival error larger than 1 second, the foreground will be extended on either side to account for this error.  The data surrounding the time of the GRB are used for background estimation, in order to assess the data quality in the detectors around the time of the GRB. 

The LIGO-Virgo detector network \cite{Abbott:2007kv, Abadie:2010px, virgostatus} is sensitive to a large fraction of the sky, albeit with relatively poor localization capability (on the order of tens or hundreds of square degrees)
\cite{Fairhurst:2010is}. In much the same way as the IPN network, the GW network of detectors can reconstruct a sky position primarily through triangulation. A GW signal with an SNR large enough to be detected will have its sky location resolved by timing its arrival at the different GW detectors.  The approximate timing resolution between two detectors in the network is $\sim$0.5ms, giving a best angular resolution of around $2^\circ$.  When performing a search for a GW signal associated with a GRB, the data from the various detectors in the network is coherently added with the appropriate time delays between detectors corresponding to a given sky location.  Thus, if the IPN error box spans a large region of the sky, it is necessary to search several different sky positions for a GW signal.

The IPN GRB error boxes may differ in shape and size, with areas ranging from fractions to hundreds or even thousands of square degrees. The short GRBs we chose for analysis have either small or very narrow and elongated error boxes which will make the gravitational waves search much easier and less computing intensive.  These error boxes are tiled by a set of points spaced by $1.8^{\circ}$ in each direction.  An empirical upper limit of 100 $\deg^{2}$ square degrees was chosen for the maximum $\Delta A$ area of the error box, necessitating a maximum of about 30 independent sky points to search the error box.  Searching over more than 30 points requires significant computational cost and much of the sensitivity improvement for the GRB triggered search over the all sky searches that have been performed would be lost.

The short GRBs for which GW data from only the H1 and H2 detectors is available are a special case.  Since these two detectors are co-located and co-alligned, they would observe any gravitational wave signal at the same time, irrespective of the location of the signal. Hence, all the search points are degenerate, i.e. using just these two detectors would not allow us any spatial sensitivity since there is no time delay between these and triangulation is impossible. A limited size error box is not a requirement for these bursts any more and any short burst, no matter how extended the error box, as long as it has available data from H1 and H2, will be analysed. Furthermore, although all-time all-sky searches for GW during S5/VSR1 have been published \cite{Abadie:2010yba, Abbott:2009dk}, these searches did not make use of the H1-H2 data in the way we will do.  This was because, as the detectors are co-located, they share many common sources of noise.  Consequently, the usual method of estimating background by introducing an artificial time shift between the detector data is not applicable.  This renders an all time search difficult as we have no accurate way of measuring the noise background.  For the previous searches only the few loudest H1-H2 coincident events were considered based on no background estimations and solely on the coincidence test. For a GRB search, we can make use of data away from the time of the GRB (without time shifting) to estimate the background, therefore providing us with a significant increase in sensitivity.

Depending on detector data availability and error box size we divide the GRB sample to be analysed into two groups: 14 short bursts with error boxes smaller than $100~\mathrm{deg}^2$ and available data from at least two sites and 6 short bursts with an arbitrarily sized error box area that have available data from H1H2 only. Six bursts have large error boxes and will be considered only for an archival look-up in the S5/VSR1 all-sky all-time data and three other bursts have already been analysed and published previously. This data is summarised in Table \ref{tab:ipn_grb}.

\begin{table}
\begin{center}
\lineup
1. Short IPN GRBs with $\Delta A < 100 ~\mathrm{deg}^2$ and data from two or more GW detector sites that will be analysed
\begin{tabular}{*{7}{l}}
\br
GRB&IPN&GW&GRB Date&$T_{90}$(s)&$\Delta A$($\mathrm{deg}^2$)&$\Delta t$(s)\cr
\br
060103&MO/I &H1H2L1&Jan 03 2006 08:42:17& 2.00&9.30&$<$1\cr
060107&K/MO/S&H1H2L1&Jan 07 2006 01:54:40&3.00&8.20&1.0\cr
060203&K/MO/H&H1H2L1&Feb 03 2006 07:28:58&0.40&0.80&$<$1\cr 
060415B&K/MO/S&H1H2L1&Apr 15 2006 18:14:44&0.44&0.20&$<$1\cr
060522C&S/K/MO&H1H2L1&May 22 2006 10:10:19&1.10&0.40&$<$1\cr 
060708B&H/K/MO&H1H2L1&Jul 08 2006 04:30:38&1.00&0.06&$<$1\cr
060930A&K/MO&H1H2L1&Sept 30 2006 02:30:11&1.00&2.40&4.5\cr
061006A&K/MO/S&H1H2L1&Oct 06 2006 08:43:38&1.60&3.20&1.0\cr
070321&I/K/MO/S&H1H2L1&Mar 21 2007 18:52:15&0.34&0.40&$<$1\cr
070414&S/M&H1H2L1&Apr 14 2007 17:19:52&0.38&0.30&$<$1\cr
070516&I/K/M/S&H1H2L1&May 16 2007 20:41:24&1.00&7.68&1.0\cr
070614&K/H&H1H2L1V1&Jun 14 2007 05:05:09&0.40&$\sim$68&$<$1\cr 
070915&Sw/I/M/K&H1H2L1V1&Sept 15 2007 08:34:48&0.50&0.10&$<$1\cr
070927A&Sw/M/I&L1V1&Sept 27 2007 16:27:55&0.70&1.60&$<$1\cr
\br
\end{tabular}

\vspace{1mm}

2. Short IPN GRBs with data from H1H2-only that will be analysed
\begin{tabular}{*{7}{l}}
\br                              
GRB&IPN&GW&GRB Date&$T_{90}$(s)&$\Delta A$($\mathrm{deg}^2$)&$\Delta t$(s)\cr
\br
060317&K/S/I &H1H2&Mar 17 2006 11:17:39&0.70&9.24&$<$1\cr
060601B&I/S&H1H2&Jun 01 2006 07:55:41&0.50&$\sim$600&$<$1\cr
061001&I/Sw&H1H2&Oct 01 2006 21:14:28&1.00&$\sim$2000&$<$1\cr
070129B&S/K&H1H2&Jan 29 2007 22:09:26&0.22&47.50&$<$1\cr
070222&K/MO&H1H2&Feb 22 2007 07:31:55&1.00&0.45&5.0\cr
070413&I/S&H1H2&Apr 13 2007 20:37:55&0.19&$\sim$350&$<$1\cr
\br
\end{tabular}

\vspace{1mm}

3. Short IPN GRBs with $\Delta A > 100 ~\mathrm{deg}^2$ and data from two or more GW detector sites for archival data look-up only
\begin{tabular}{*{7}{l}}
\br
GRB&IPN&GW&GRB Date&$T_{90}$(s)&$\Delta A$($\mathrm{deg}^2$)&$\Delta t$(s)\cr
\br
060916&S/I&H1H2L1&Sept 16 2006 14:33:34&0.13&$>$3000&$<$1\cr
061014&I/H&H1L1&Oct 14 2006 06:17:02&1.5&$>$3000&$<$1\cr
061111B&K/Sw&H1H2L1&Nov 11 2006 10:54:27&0.6&$\sim$700&$<$1\cr
070203&I/S&H1H2L1&Feb 03 2007 23:06:44&0.69&$>$2000&$<$1\cr
070721C&K/I&H1H2V1&Jul 21 2007 14:24:09&1.00&495&$<$1\cr
070910&K/S&H1H2L1V1&Sept 10 2007 17:33:29&0.38&$>$200&$<$1\cr
\br
\end{tabular}

\vspace{2mm}

4. Short IPN GRBs that have already been analysed and published
\begin{tabular}{*{3}{l}}
\br
GRB&IPN&Reference\cr
\br
060427&K/MO/I/Sw&\cite{Abadie:2010uf, Collaboration:2009kk}\cr
060429A&S/K/MO&\cite{Abadie:2010uf, Collaboration:2009kk}\cr
070201&K/M/I&\cite{Abbott:2007rh}\cr
\br
\end{tabular}

\vspace{2mm}

\caption{\label{tab:ipn_grb}The short S5/VSR1 IPN GRB sample - 14 with data from multiple non-H1H2-only GW detectors and well localised bursts (error box area $\Delta A <$100$~\mathrm{deg}^2$);  6 H1H2-only poorly localised bursts; 6 multiple GW detectors for poorly localised bursts; 3 bursts previously analysed and published. $\Delta t$ represents the time of arrival error. The IPN satellites that observed the bursts: S - Suzaku, Sw - Swift, I - INTEGRAL, M - MESSENGER, MO - Mars Odyssey, K - Konus-WIND, H - HESSI (RHESSI).}
\end{center}
\end{table}

\section{The proposed search}

For the IPN short GRBs, the data streams from the operational detectors will be combined coherently and searched using the methods described in \cite{Harry:2010fr}. The search for compact binary coalescing signals is done using match filtering \cite{OwenSathyaprakash98} by correlating the detector data against theoretical waveforms that replicate the signal for a broad interval of binary parameters. In order to cope with the affects of non-stationary, transient noise ``glitches'' in the GW detectors' data, the pipeline uses a number of signal consistency tests, including the null stream, amplitude consistency and several $\chi^{2}$ tests \cite{Allen:2004gu, Harry:2010fr, Hanna:2008}. The pipeline can do the search on a single sky point as well as on multiple sky points. Simulated GW signals injected in the data streams are used to assess the sensitivity of the search. The two most important changes specific to the analysis of the IPN GRBs are the way we will perform the search over multiple sky points and the generation of simulations over the GRB error boxes.

We will generate a grid of search points to pave each GRB's error box in order to increase the chances of finding a signal over the entirety of the box. A fixed grid spacing (distance between adjacent grid points) of 1.8 degrees will be used for the search, motivated by the GW detectors' power of resolving the sky location, described above. Each error box is a 3-$\sigma$ region but we will search and assign equal detection probability to each search point.

Aside from generating a search grid, we will also generate random positions in right ascension and declination within an IPN error box for simulations. To generate these positions, the error box is first paved with grid points with a denser 0.2 degree grid spacing. This is a finer spacing than the one which is used for the search to ensure efficient coverage of the error box by the simulated positions. It will also provide a verification that the 1.8 degree spacing of search points is adequate.  Random positions are generated within a small square bin centred at each of these grid points and whose sides have lengths equal to the grid spacing. The relative number (or density) of positions generated for each grid point is weighted according to the estimated source position probability distribution. The probability distribution that will be used in the case of a single 3-$\sigma$ IPN annulus is a one-dimensional Gaussian distribution centred at the central radius of the IPN annulus, and which has a sigma of 1/3 the width of the given annulus half-width. For error boxes which are formed by the intersection of two 3-$\sigma$ IPN annuli, the probability distribution will be a two-dimensional Gaussian, with distances measured from the two central radii of the two annuli. This assures that there are proportionally more simulations for those positions with larger probabilities of having a signal. Then we will draw random locations for injections from this list of simulation points.

\section{Discussion}

We have presented the methodology for a search for gravitational waves around the times of short GRBs, detected by IPN during S5 and VSR1. This search has all the needed tools and will commence in a short time. The work in this paper was presented in the form of a poster at the 10th Amaldi Conference for Gravitational Waves organised in Cardiff, UK in July 2011. These types of searches are very promising for the future detection of gravitational waves and combined with the prospect of detecting other electromagnetic counterparts from GRBs, e.g. radio or optical pre- or afterglows (summarised in \cite{Predoi:2009af, Coward:2011yr}), may open the doors for true multi-messenger astronomy with gravitational waves.

\section*{Acknowledgements}

The authors gratefully acknowledge the support of the United States
National Science Foundation for the construction and operation of the
LIGO Laboratory, the Science and Technology Facilities Council of the
United Kingdom, the Max-Planck-Society, and the State of
Niedersachsen/Germany for support of the construction and operation of
the GEO600 detector, and the Italian Istituto Nazionale di Fisica
Nucleare and the French Centre National de la Recherche Scientifique
for the construction and operation of the Virgo detector. The authors
also gratefully acknowledge the support of the research by these
agencies and by the Australian Research Council,
the International Science Linkages program of the Commonwealth of Australia,
the Council of Scientific and Industrial Research of India,
the Istituto Nazionale di Fisica Nucleare of Italy,
the Spanish Ministerio de Educaci\'on y Ciencia,
the Conselleria d'Economia Hisenda i Innovaci\'o of the
Govern de les Illes Balears, the Foundation for Fundamental Research
on Matter supported by the Netherlands Organisation for Scientific Research,
the Polish Ministry of Science and Higher Education, the FOCUS
Programme of Foundation for Polish Science,
the Royal Society, the Scottish Funding Council, the
Scottish Universities Physics Alliance, The National Aeronautics and
Space Administration, the Carnegie Trust, the Leverhulme Trust, the
David and Lucile Packard Foundation, the Research Corporation, and
the Alfred P. Sloan Foundation. Kevin Hurley is grateful for IPN support from
the following NASA grants: NNX10AR12G (Suzaku), NNX07AR71G (MESSENGER),
NNX10AI23G (Swift), NNX09AR28G (INTEGRAL), and JPL 1282043 (Odyssey).
 
\section*{References}
\bibliographystyle{unsrt}
\bibliography{references}

\begin{thebibliography}{10}

\bibitem{Nakar:2007}
Ehud Nakar.
\newblock {Short-hard gamma-ray bursts}.
\newblock {\em Phys. Rept.}, 442:166--236, 2007.

\bibitem{ShibTan06}
Masaru Shibata and Keisuke Taniguchi.
\newblock Merger of binary neutron stars to a black hole: Disk mass, short
  gamma-ray bursts, and quasinormal mode ringing.
\newblock {\em Phys. Rev.}, D73:064027, 2006.

\bibitem{Berger:2010qx}
Edo Berger.
\newblock {The Environments of Short-Duration Gamma-Ray Bursts and Implications
  for their Progenitors}.
\newblock {\em New Astron. Rev.}, 55:1--22, 2011.

\bibitem{ACST94}
Theocharis~A. Apostolatos, Curt Cutler, Gerald~J. Sussman, and Kip~S. Thorne.
\newblock Spin induced orbital precession and its modulation of the
  gravitational wave forms from merging binaries.
\newblock {\em Phys. Rev.~D}, 49:6274--6297, 1994.

\bibitem{Kiuchi:2010ze}
Kenta Kiuchi, Yuichiro Sekiguchi, Masaru Shibata, and Keisuke Taniguchi.
\newblock {Exploring binary-neutron-star-merger scenario of short- gamma-ray
  bursts by gravitational-wave observation}.
\newblock {\em Phys. Rev. Lett.}, 104:141101, 2010.

\bibitem{Abadie:2010uf}
J.~Abadie et~al.
\newblock {Search for gravitational-wave inspiral signals associated with short
  Gamma-Ray Bursts during LIGO's fifth and Virgo's first science run}.
\newblock {\em Astrophys. J.}, 715:1453--1461, 2010.

\bibitem{Collaboration:2009kk}
B.~P. Abbott et~al.
\newblock {Search for gravitational-wave bursts associated with gamma-ray
  bursts using data from LIGO Science Run 5 and Virgo Science Run 1}.
\newblock {\em Astrophys. J.}, 715:1438--1452, 2010.

\bibitem{Hurley:2002wv}
K.~Hurley et~al.
\newblock {The Current Performance of the Third Interplanetary Network}.
\newblock {\em AIP Conf. Proc.}, 662:473--476, 2003.

\bibitem{HurleyHTML}
http://www.ssl.berkeley.edu/ipn3/chronological.csv.

\bibitem{Hurley:1999ym}
K.~Hurley et~al.
\newblock {Precise Interplanetary Network Localization of the Bursting Pulsar
  GRO J1744-28}.
\newblock 1999.

\bibitem{lvc:s6grb}


\bibitem{Harry:2010fr}
Ian~W. Harry and Stephen Fairhurst.
\newblock {A targeted coherent search for gravitational waves from compact
  binary coalescences}.
\newblock {\em Phys. Rev.}, D83:084002, 2011.

\bibitem{Qinx:2010kp}
Y.~P. Qinx, A.~C. Gupta, J.~H. Fan, C.~Y. Su, and R.~J. Lu.
\newblock {Duration distributions for different softness groups of gamma-ray
  bursts}.
\newblock {\em Sci. China Phys. Mech. Astron.}, 53:1375--1382, 2010.

\bibitem{McBreen:2001fd}
S.~McBreen, F.~Quilligan, B.~McBreen, L.~Hanlon, and D.~Watson.
\newblock {Temporal properties of the short gamma-ray bursts}.
\newblock 2001.

\bibitem{Blanchet:2001aw}
Luc Blanchet, Bala~R. Iyer, and Benoit Joguet.
\newblock {Gravitational waves from inspiralling compact binaries: Energy flux
  to third post-Newtonian order}.
\newblock {\em Phys. Rev.}, D65:064005, 2002.

\bibitem{BD89}
L.~Blanchet and T.~Damour.
\newblock Post-newtonian generation of gravitational waves.
\newblock {\em Annales Inst. H. Poincar\'e Phys. Th\'eor.}, 50:377--408, 1989.

\bibitem{Shibata:2005mz}
Masaru Shibata, Matthew~D. Duez, Yuk~Tung Liu, Stuart~L. Shapiro, and
  Branson~C. Stephens.
\newblock {Magnetized hypermassive neutron star collapse: a central engine for
  short gamma-ray bursts}.
\newblock {\em Phys. Rev. Lett.}, 96:031102, 2006.

\bibitem{Duez:2005cj}
Matthew~D. Duez, Yuk~Tung Liu, Stuart~L. Shapiro, Masaru Shibata, and
  Branson~C. Stephens.
\newblock {Collapse of magnetized hypermassive neutron stars in general
  relativity}.
\newblock {\em Phys. Rev. Lett.}, 96:031101, 2006.

\bibitem{Shibata:2007zm}
Masaru Shibata and Keisuke Taniguchi.
\newblock {Merger of black hole and neutron star in general relativity: Tidal
  disruption, torus mass, and gravitational waves}.
\newblock {\em Phys. Rev.}, D77:084015, 2008.

\bibitem{Rezzolla:2011da}
Luciano Rezzolla et~al.
\newblock {The missing link: Merging neutron stars naturally produce jet-like
  structures and can power short Gamma-Ray Bursts}.
\newblock {\em Astrophys. J.}, 732:L6, 2011.

\bibitem{Oechslin:2005mw}
R.~Oechslin and Hans-Thomas Janka.
\newblock {Torus Formation in Neutron Star Mergers and Well-Localized Short
  Gamma-Ray Bursts}.
\newblock {\em Mon. Not. Roy. Astron. Soc.}, 368:1489--1499, 2006.

\bibitem{ABIQ04}
K~G Arun, Luc Blanchet, Bala~R. Iyer, and Moh'd S~S. Qusailah.
\newblock The 2.5pn gravitational wave polarisations from inspiralling compact
  binaries in circular orbits.
\newblock {\em Class. Quantum Grav.}, 21:3771, 2004.
\newblock Erratum-ibid. {\bf 22}, 3115 (2005).

\bibitem{AIQS06a}
K~G Arun, B~R Iyer, M~S~S Qusailah, and B~S Sathyaprakash.
\newblock Testing post-newtonian theory with gravitational wave observations.
\newblock {\em Class. Quantum Grav.}, 23:L37, 2006.

\bibitem{heasarc}
http://heasarc.gsfc.nasa.gov/W3Browse/all/ipngrb.html.

\bibitem{suzaku}
http://www.astro.isas.ac.jp/suzaku/HXD WAM/WAM-GRB/.

\bibitem{integral}
http://www.isdc.unige.ch/integral/science/grb.

\bibitem{swift}
http://swift.gsfc.nasa.gov/docs/swift/swiftsc.html.

\bibitem{konus}
http://lheawww.gsfc.nasa.gov/docs/gamcosray/legr/bacodine/konus$\_$grbs.html.

\bibitem{Abbott:2007kv}
B.~Abbott et~al.
\newblock {LIGO: The Laser Interferometer Gravitational-Wave Observatory}.
\newblock {\em Rept. Prog. Phys.}, 72:076901, 2009.

\bibitem{Abadie:2010px}
J.~Abadie et~al.
\newblock {Calibration of the LIGO Gravitational Wave Detectors in the Fifth
  Science Run}.
\newblock {\em Nucl. Instrum. Meth.}, A624:223--240, 2010.

\bibitem{virgostatus}
Virgo Collaboration.
\newblock Virgo status.
\newblock {\em Classical and Quantum Gravity}, 25(18):184001, 2008.

\bibitem{Fairhurst:2010is}
Stephen Fairhurst.
\newblock {Source localization with an advanced gravitational wave detector
  network}.
\newblock {\em Class. Quant. Grav.}, 28:105021, 2011.

\bibitem{Abadie:2010yba}
J.~Abadie et~al.
\newblock {Search for Gravitational Waves from Compact Binary Coalescence in
  LIGO and Virgo Data from S5 and VSR1}.
\newblock {\em Phys. Rev.}, D82:102001, 2010.

\bibitem{Abbott:2009dk}
B.~Abbott et~al.
\newblock Search for gravitational waves from low mass binary coalescences in
  the first year of ligo's s5 data.
\newblock 2009.

\bibitem{Abbott:2007rh}
B.~Abbott et~al.
\newblock {Implications for the Origin of GRB 070201 from LIGO Observations}.
\newblock {\em Astrophys. J.}, 681:1419--1428, 2008.

\bibitem{OwenSathyaprakash98}
Benjamin~J. Owen and B.~S. Sathyaprakash.
\newblock {Matched filtering of gravitational waves from inspiraling compact
  binaries: Computational cost and template placement}.
\newblock {\em Phys. Rev.}, D60:022002, 1999.

\bibitem{Allen:2004gu}
Bruce Allen.
\newblock {A chi**2 time-frequency discriminator for gravitational wave
  detection}.
\newblock {\em Phys. Rev.}, D71:062001, 2005.

\bibitem{Hanna:2008}
Chad. Hanna.
\newblock {\em Searching for gravitational waves from binary systems in
  non-stationary data}.
\newblock PhD thesis, Louisiana State University, 2008.

\bibitem{Predoi:2009af}
V.~Predoi et~al.
\newblock {Prospects for joint radio telescope and gravitational wave searches
  for astrophysical transients}.
\newblock {\em Class. Quant. Grav.}, 27:084018, 2010.

\bibitem{Coward:2011yr}
D.~M. Coward et~al.
\newblock {Toward an optimal search strategy of optical and gravitational wave
  emissions from binary neutron star coalescence}.
\newblock {\em Mon. Not. Roy. Astron. Soc.}, 415:L26, 2011.

\end{thebibliography}
\end{document}